# PLOS/Mozilla Scientific Code Review Pilot: Summary of Findings

Prof. Marian Petre (Open University) and Dr. Greg Wilson (Mozilla Foundation)
November 2013

*PLOS and Mozilla conducted a month-long pilot study in which professional developers performed code reviews on software associated with papers published in PLOS Computational Biology. While the developers felt the reviews were limited by (a) lack of familiarity with the domain and (b) lack of two-way contact with authors, the scientists appreciated the reviews, and both sides were enthusiastic about repeating the experiment.*

**Background**

Since Fagan's pioneering work at IBM in the mid 1970s [1,2], dozens of studies have shown that code review is the most effective to find bugs in software [3]. It is also the most time-effective: hour for hour, nothing is better at improving code quality than having someone other than the author read it carefully.

Despite these findings, code review was still talked about more than done until the early 2000s. What changed was the rise of open, distributed collaboration, particularly in open source projects that didn't have the command-and-control structures of commercial software shops. Today, code review is routine in every large open source project. Programmers don't make changes directly to the project's master copy; instead, they post changes for review, read and comment on each other's changes, make fixes where needed, and only then commit those improvements.

Ironically, given that Fagan and others were inspired by academic peer review, code review is still rare in scientific software projects. This is partly a case of the blind leading the blind (faculty don't do code reviews, so graduate students don't learn the practice, so they don't pass it on to their students), and partly because most authors don't publish code with their papers, and hence have little incentive to improve its quality or readability.

The fundamental cause, though, is that many scientists do not understand the real benefits of code review. Its goal is not to find bugs once the code is "done", but to reduce the time required to produce correct code by reducing the re-work needed to fix bugs weeks or months after they have been introduced. In a phrase, "measure twice, cut once" speeds up programming just as much as it speeds up carpentry, and for the same reasons.

In August and September 2013, PLOS and Mozilla Science Lab ran a small pilot study to explore code review of scientific code. In this study, professional software developers working at Mozilla reviewed samples of code taken from papers published in PLOS Computational Biology in the preceding 18 months. Those reviews were then shared with the scientists, and both groups were interviewed to explore:

- whether non-scientists could usefully review typical scientific software;
- whether those reviews were intelligible and useful to the scientists; and
- whether the participants felt the reviews were valuable.

We interviewed 11 developers (9 of whom had completed code reviews) in late August and early September, and 4 of the code authors whose code had been reviewed (3 scientists, one developer on a team of scientists) in late September and early October. The interviews were semi-structured, with questions to elicit both the reviewers' and the scientists' perceptions of the conduct, value and utility of the pilot reviews. Semi-structured interviews are a standard approach for exploratory studies [4, 5], because they provide a basis for comparability across interviews while permitting flexibility in following up areas of interest with interviewees. Interviews were conducted via Skype or telephone and typically lasted 20 to 30 minutes. Detailed contemporaneous notes were taken, and interviews were audio-recorded subject to the permission of the interviewee.

Reviewers were asked to compare the review of the scientific code with their previous professional experience of code reviews. They were asked about their ability to engage with the scientific context, and if they felt they could offer useful commentary. Authors were asked if code review would be of value in their future practice. All were asked about their technical background, prior experience with code review, and whether they would be willing to participate in a further pilot study.

The interviews were analysed in terms of the questions that drove the pilot study. Responses were tabulated under the key elements embodied in the interview framework, and overall patterns in the responses were characterized. This interim report summarises the major themes emerging from those interviews.

1. **The scientists who agreed to be interviewed likely place more value on code quality than the average scientist whose work involves code.**

Although the scientists described themselves as largely self-taught programmers, they could be characterized as using some engineering practices and tools, some of the time. For example, most of the code authors interviewed use open software repositories, but not all contribute their code back into the repository; the use version control, but only when then code gets big); they may use unit testing. Documentation is rare: as we discuss below, there is less commenting (both internal and external) than developers are used to, and practices like test-driven development were unfamiliar or considered "very unusual".

Not all of the scientists were familiar with code review. Some of the code authors we interviewed work in (or lead) teams in which they read each other's code. Others "almost never" get to discuss their code with others: they discuss results, not how the code produced them. One, who does a weekly code review with his team, observed that their "in-code documentation" has improved through code review. He has a checklist for code review in his lab, and asked if the Mozilla developers would review it to consider whether there are things that should be added.

2. **The reviewers felt the reviews were shallow, but the scientists appreciated them.**

Overall, all of the reviewers were satisfied that they were able to provide a review with some utility, but they were also frustrated because their review was constrained by lack of domain knowledge, lack of communication with the author – and hence lack of context. Therefore, they felt that their reviews were "limited" or "shallow". A number of reviewers noted that they were unable to run the code, which most reviewers (but not all) would normally do. As a result, the review comments were at the algorithmic, syntactic, and stylistic levels; they were not able to comment on structure (except at low level), organization or architecture.

In contrast, all of the code authors found the comments useful: typical responses were, "detailed", "knowledgeable reviewer", "presentation clear", "gave alternative suggestion", and "useful feedback". Feedback the scientists found particularly valuable focused on:

- usability (e.g., how easily someone can "enter the structure" of the code)
- ease of re-use, readability, and density
- code structure and solution structure ("The way they organise a solution to a problem")
- feedback on the organisation of README files

- performance and opttimization
- unit testing (or its absence)

From this, we may conclude that there is low-hanging fruit in scientific code review: things that do not seem challenging to software developers are nevertheless seen to be useful by scientists who are striving to introduce a culture of code quality.

3. **The developers felt constrained by their lack of domain knowledge.**

Few of the reviewers read in detail the paper with which the code was associated, usually because they felt they lacked the domain expertise to do so. (None claimed to have a scientific background appropriate for the level of the publication they handled.) Most scanned the paper or read the abstract. One commented that, having scanned the code and read the abstract, he realized that reading the paper wouldn't help him in understanding the code. One noted that the code didn't actually do the analysis with which the paper was concerned. Another reported that he sought out the larger code base from which the review code had been selected. (There were some issues with the code pulled for the pilot.)

The lack of domain expertise, combined with isolation from the scientists, meant that none of the reviewers felt able to assess whether the code implemented the science effectively. They could deduce what the code did and assess it in terms of its ability to perform that operation efficiently, but they could not determine if the code fulfilled its role in the scientific context.

4. **The developers missed being able to run the code.**

*All* of the developers doing reviews expressed some frustration over the limitations of the material they had to review. The lack of documentation, of commenting in the code, of tests, and of example data sets was noted. The omissions in the package had consequences on the review process. "Not having the project build is a big problem; I can't verify that the code is correct." Most of the reviewers normally run the code as the first step in review: "That's the easiest way to see if it works at all," and "It's a way to validate the intentions of the author." Several commented that, because they could not run the code or assess its correctness, they had to make some assumptions in the review.

One of the biggest frustrations concerned the reviewers' ignorance of the scientists' intentions, or of what sort of review would be useful for the scientists. The lack of documentation ("Basic description: what it is, what it does...") left the reviewers without a sufficient understanding of the context. Most noted that they spent more of the review on basics, because it was "not easy to form a picture of the whole code". Several likened it to a "drive-by" review.

5. **The reviewers felt the scientists had less concern for maintainability and readability than they were used to in professional code, but all of the code authors aspire to readable, re-usable code.**

Many of the reviewers were struck by how the scientific code differed from professional code, and remarked on the lack of commenting and explanation in the code. Several suggested that the scientists appear to have "less concern for maintainability and readability" and that the code was "not written for others to use". Some also pointed at what appeared to them to be novice or naive characteristics in the code, citing a lack of complexity, of use of abstraction; redundancy in the code; inconsistencies or ignorance of standards in formatting; and unhelpful naming. One said that he had expected scientists to produce code that "barely works", that is, code that works when used as intended, but that is neither robust nor readable outside the original context.

On their side, the scientists all aspire to code that is readable and re-usable, but many noted that their code doesn't respect the common etiquette of open source. For example, they often don't make an effort to package their code for re-use by others. One scientist went further and commented that: "[It's] not important to have something that's exact - only when we publish" and reiterated the low status of code in their science: "In the business of science, all that matters is the figures. The quality of the code is just not on the critical path."

This "culture gap" could be reduced (possibly eliminated) by drawing reviewers from the same scientific culture as the code's authors, or by having developers and scientists work together, in a dialogue, over a longer period (so that the former come to understand the incentives and culture of the latter). Both developers and scientists expressed a desire to learn from the other's experience and perspective.

**6. The reviewers had a strong sense of the audience for their reviews.**

Despite their isolation from the scientists, most of the reviewers had a mental image of the code author (i.e., of the recipient of their review), usually based on clues in the code itself, sometimes augmented by other experience of scientist developers or open source communities. One reviewer expressed it as using the same voice as he would for a Mozilla coder, but addressing more basic content (i.e., for a respected colleague doing an unfamiliar job). Another said: "They didn't use a very sophisticated method for putting the algorithm together: easy to interpret, but harder to reuse (...lacking abstraction)."

**7. The reviewers felt the lack of social context, and wanted to be able to have a dialog with the scientists, rather than doing a "drive by" review.**

All but one of the reviewers remarked that they miss the social context they normally associate with code reviews: working relationships, understanding goals and priorities, and trust. The social context helps them to establish a shared understanding of the particular task, and hence to provide reviews that are appropriate and useful. They would prefer some form of dialog, not least to provide them with the code context, and to establish appropriate expectations for both reviewers and authors (e.g., "knowing what kind of feedback the author wants"). One typical remark was, "It would have been easier if we were allowed to contact the scientist just to get a feel for his mindset."

Most referred to dialogs that are normally part of their code review experience, whether spoken or on-line: "Discussion catches details that get lost otherwise" such as "tiny changes in numerical interpretation that are important". Without interaction between reviewer and author, we "lose a lot of discussion about how you might want to think about your process". Finally, the reviewers pointed out that the dialog has value for the reviewer, as well as for the author. "When there is a dialog, you end up learning a lot yourself."

**8. The reviewers would do it again, especially if there were some form of dialogue with the author.**

Perhaps the most positive outcome of the pilot was the reviewers' willingness to continue to engage in scientific code review, provided they felt they could contribute something of value. Nearly every reviewer wanted to know if their reviews are useful to the authors – or if they would be in a less artificial, more interactive, context in which they could interact directly with the scientists.

Several made proposals about developing a communication space (akin to practice in OS), in which scientists could solicit code reviews and interact with the reviewers. They asked for "...other things that go with code review: mailing lists, IRC, free-flowing interaction, discussion group...", i.e., for a means to communicate and share ideas. A typical comment was: "You need a place for researchers to go where they can ask...that's the real benefit: share ideas and get better."

Many see code review as a form of professional development, and they value their role as mentors: "The thing that's missing: mentoring, guidance about how to write code that will turn out well in review..." They emphasise that code review is most effective *during* the development process; they considered that conducting code review after the research was published was "... too late to learn something".

### 9. The scientists were equally positive.

All of the scientific code authors we interviewed were positive about the pilot, making comments such as: "Great idea...the implementation is important in how well it will work." *All* volunteered to participate in a Phase 2 that involved providing code of their choice for comment as part of a developmental process. They said that they would welcome access to professional code review.

Two of the scientists interviewed showed an interest in validation: "If the code would be validated by the Mozilla group, that would be useful." Another felt that submitting code at the same time as the article would have value if an expert developer would say "that's good for sharing"; as one put it, "[A] friendly policemen would be a good idea." Tellingly, one scientist asked, "How [to] get a funding agency to pay for it?" It hadn't occurred to him that anyone would offer code review for free, despite the fact that peer review of scientific papers is almost always done *gratis*.

**Conclusion**

In summary, the participants were enthusiastic about the pilot despite the technical hiccups. All appreciated the potential of code review to improve both the quality of science based on computation, and the re-usability of the scientific code itself. All of those interviewed volunteered to participate in a second phase of the pilot study, and were interested in exposing their code to professional review during its development, as part of a dialogue. In contrast to the concerns expressed by biostatistician Roger Peng at the Johns Hopkins Bloomberg School of Public Health, as quoted in a recent *Nature* report [6] that "...with reviews like this, scientists will be even more discouraged from publishing their code", scientists welcomed the opportunity to improve their code.